\documentclass[aps,prl,reprint,floatfix,citeautoscript,longbibliography,superscriptaddress]{revtex4-1}
\usepackage{graphicx}
\usepackage{bm,amsmath,amssymb,mathrsfs,dcolumn}
\usepackage{gensymb}
\usepackage[usenames]{color}
\usepackage{subfigure}
\usepackage{ulem}
\usepackage{hyperref}
\hypersetup{colorlinks=true, linkcolor=blue, citecolor=red, urlcolor=blue}
\begin{document}

\title{Steady Bell state generation via magnon-photon coupling}
\author{H. Y. Yuan}
\email[Electronic address: ]{yuanhy@sustech.edu.cn}
\affiliation{Department of Physics, Southern University of Science and Technology, Shenzhen 518055, China}
\author{Peng Yan}
\affiliation{School of Electronic Science and Engineering and State Key Laboratory of
Thin Films and Integrated Devices, University of Electronic Science and Technology of China,
Chengdu 610054, China}
\author{Shasha Zheng}
\affiliation{State Key Laboratory for Mesoscopic Physics, School of Physics $\&$ Collaborative Innovation Center of Quantum Matter, Peking University, Beijing 100871, China}
\affiliation{Beijing Academy of Quantum Information Sciences, Haidian District, Beijing 100193, China}
\affiliation{Collaborative Innovation Center of Extreme Optics, Shanxi University, Taiyuan, Shanxi 030006, China}
\author{Q. Y. He}
\affiliation{State Key Laboratory for Mesoscopic Physics, School of Physics $\&$ Collaborative Innovation Center of Quantum Matter, Peking University, Beijing 100871, China}
\affiliation{Beijing Academy of Quantum Information Sciences, Haidian District, Beijing 100193, China}
\affiliation{Collaborative Innovation Center of Extreme Optics, Shanxi University, Taiyuan, Shanxi 030006, China}
\author{Ke Xia}
\affiliation{Institute for Quantum Science and Engineering and Department of Physics, Southern University of Science and Technology, Shenzhen, 518055, China}
\author{Man-Hong Yung}
\email[Electronic address: ]{yung@sustech.edu.cn}
\affiliation{Institute for Quantum Science and Engineering and Department of Physics, Southern University of Science and Technology, Shenzhen, 518055, China}
\affiliation{Shenzhen Key Laboratory of Quantum Science and Engineering, Shenzhen 518055, China}
\date{\today}

\begin{abstract}
We show that parity-time ($\mathcal{PT}$) symmetry can be spontaneously broken
in the recently reported energy level attraction of magnons and cavity photons. In the $\mathcal{PT}$-broken phase,
magnon and photon form a high-fidelity Bell state with maximum entanglement.
This entanglement is steady and robust against the perturbation of environment, in contrast to the general wisdom that expects instability of the hybridized state when the symmetry is broken. This anomaly is further understood by the compete of non-Hermitian evolution and particle number conservation of the hybridized system.
As a comparison, neither $\mathcal{PT}$-symmetry broken nor steady magnon-photon entanglement is observed inside the normal level repulsion case. Our results may open a novel window to utilize magnon-photon entanglement as a resource for quantum technologies.
\end{abstract}

\maketitle
{\it Introduction.---} Manipulating light-matter interaction is a long lasting and intriguing topic for
its pivotal role in quantum science and technologies. Recently, the strong coupling of magnons and cavity photons was intensively investigated with the aim of realizing quantum information transfer in hybridized spintronic systems \cite{Soykal2010,Tabuchi2014, Zhang2014, Gor2014,Huebl2013, Cao2015, Wang2018,Vahram2018,Harder2018, Bohi2019}.
Historically, the coherent magnon-photon coupling
 with a typical energy level repulsion (anticrossing) spectrum was first identified
by placing magnetic insulator yttrium-iron-garnet (YIG) into a microwave cavity and/or coplanar waveguide \cite{Soykal2010,Tabuchi2014, Zhang2014, Gor2014,Huebl2013, Cao2015, Wang2018}, while recent theory and experiment show that an abnormal anticrossing (energy level attraction) spectrum emerges by considering the feedback effect of photons \cite{Vahram2018,Harder2018, Bohi2019}. Near the energy level repulsion, the magnon and photon hybridize
to form an effective two-level platform, which launches Rabi oscillation of the polariton and enables the coherent information transfer between magnons and photons \cite{Tabuchi2015}. However, the two energy levels of magnons and photons merge into a single band in the level attraction case and it is not known how magnons and photons interplay inside the band. This issue is urgent if one tends to bridge cavity spintronics with quantum information science, in which entanglement is an indispensable resource. Furthermore, the magnon-photon system with feedback effect is not Hermitian any more and
there may exist complex eigenmodes \cite{Vahram2018,Harder2018, Bohi2019}. This intriguing feature provides a generic platform to study non-Hermitian quantum physics and $\mathcal{PT}$-symmetry \cite{Gan2018, yan2019}.

In this work, we study the quantum correlation of magnons and photons inside the level attraction regime by solving the non-Hermitian dynamic equation and find that magnon and photon form a maximally entangled Bell state in the $\mathcal{PT}$-broken phase of the system. This Bell state
is steady that does not decay with time. Compared with the traditional methods of generating Bell states \cite{Kwiat1995, Kow2012,Shwa2013}, our proposal is of high fidelity, deterministic, and robust against dissipation. As we tune the magnon frequency, the system undergoes a phase transition to
the $\mathcal{PT}$-exact phase, the steady entanglement is replaced by an oscillating entanglement. Our results may open the door of non-Hermitian spintronics with $\mathcal{PT}$ symmetry and it also provides a new route to use the the entangled magnon polariton as an entanglement resource.


{\it General formalism.---} We consider a magnon-photon hybridized system with feedback action of photons,
which has been realized by placing an magnet on an inverted pattern of split-ring resonator or into a Fabry-Perot-like cavity \cite{Harder2018, Bohi2019}.
The Hamiltonian of such a hybridized system can be written as
$\mathcal{H}=\mathcal{H}_{\mathrm{FM}}
+1/2\int \left ( \epsilon_0 \mathbf{E}^2
+ \mathbf{B}^2 /\mu_0 \right ) dV -\sum_i \mathbf{S}_i \cdot \mathbf{h}_f$,
where the first, second, and third terms are the
ferromagnetic, electromagnetic (EM) wave, and interaction parts of the Hamiltonian, respectively. $\mathcal{H}_{\mathrm{FM}}$ includes the exchange, anisotropy and Zeeman energy. $\mathbf{E}$ and $\mathbf{B}$
 are respectively the electric and magnetic components of the EM wave while $\epsilon_0$ and
$\mu_0$ are vacuum permittivity and susceptibility, respectively. $\mathbf{S}_i$ is the spin of $i-$th site while the oscillating field $\mathbf{h}_f$ acting on the local spin includes a direct action of microwave $\mathbf{h}_1=\mathbf{h}e^{-i\omega_c t}$  and a reaction field of the precessing magnetization $\mathbf{h}_2= \mathbf{h}_1 \delta e^{i\phi}$ \cite{Harder2018, Bohi2019}, where $\omega_c$ is the microwave frequency, $\delta$ and $\phi$ are respectively the relative amplitude and phase of the two waves.
In the low energy limit, we follow the standard quantization procedures of magnons and photons \cite{yuan2018apl,Yuan2019}, and write the Hamiltonian as,
 \begin{equation}
\begin{aligned}
\mathcal{H}&=  \omega_r a^\dagger a + \omega_c c^\dagger c  + g \left ( a^ \dagger c + e^{i\Phi} a c^\dagger \right),
\end{aligned}
\label{ham}
\end{equation}
where $a,c,a^\dagger, c^\dagger$ are annihilation and creation operators for magnons and photons, respectively.
$\omega_r$ is the magnon frequency, $g$ is the effective coupling strength
of magnons and photons, and $\tan \Phi/2 =-\delta \sin \phi/(1+\delta \cos \phi)$ is a tunable phase factor coming from the backaction effect. The effective Hamiltonian (\ref{ham}) can describe the dissipative magnon-photon coupling in
the experimental setup \cite{Harder2018}.

When $\Phi = k \pi~ (k=0,1,2...)$, it is straightforward to show that the $\mathcal{PT}$ operation commutes with the Hamiltonian,
such that the system respects $\mathcal{PT}$-symmetry. However, this does not guarantee that the $\mathcal{PT}$ operator and Hamiltonian
display the same set of eigenstates due to the anti-linearity of $\mathcal{PT}$ operator \cite{Sch2011,Bender1998}.
If $\mathcal{PT}$ and $\mathcal{H}$ share simultaneous eigenstates with real eigenvalues, the phase is denoted as $\mathcal{PT}$-exact phase. Otherwise, the phase is $\mathcal{PT}$-broken characterized by complex eigenvalues \cite{Gan2018}.

\begin{figure}
  \centering
  \includegraphics[width=0.4\textwidth]{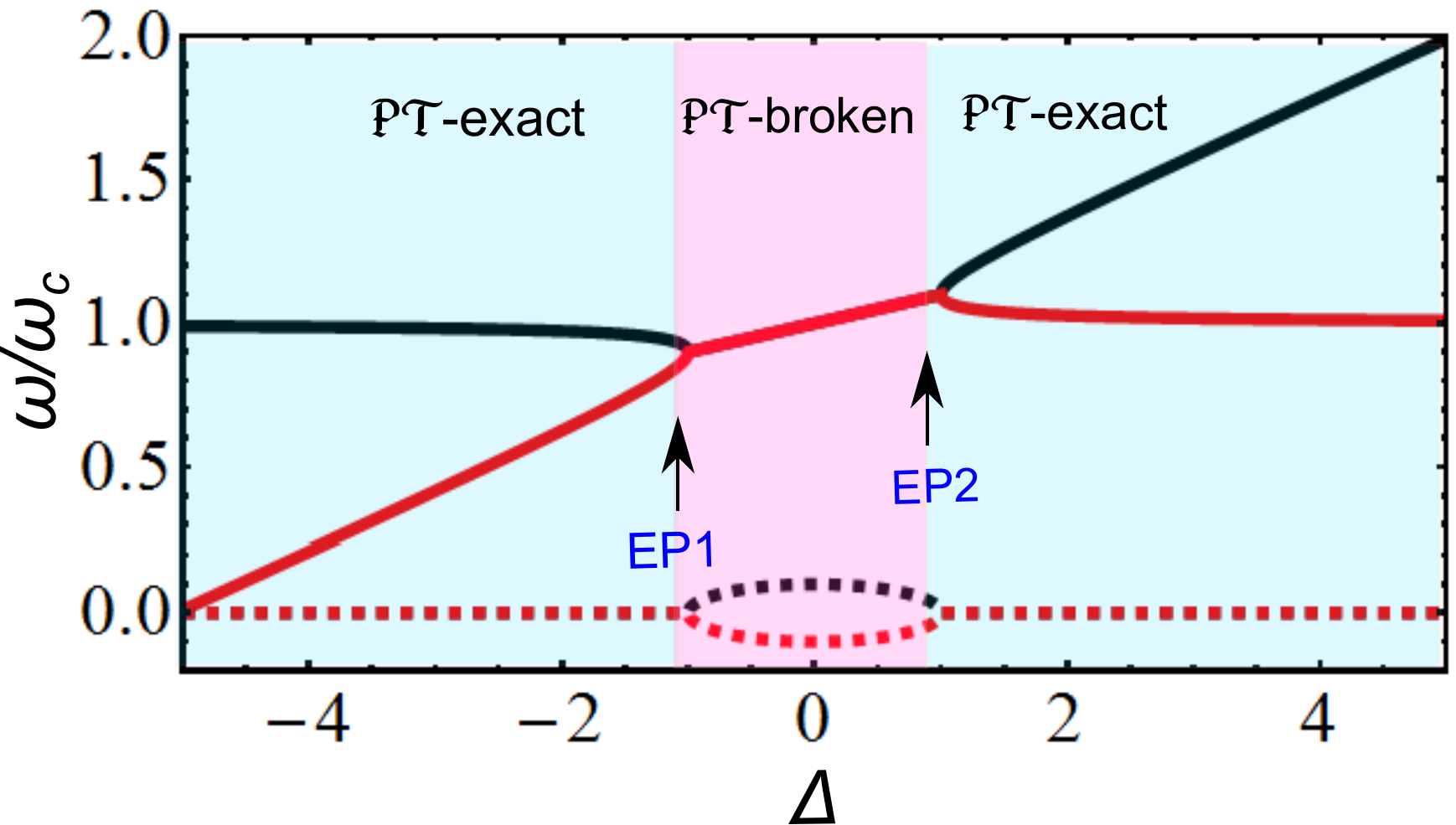}\\
  \caption{ Level attraction of the system described by Eq. (\ref{ham}) for $\Phi=\pi$. The solid and dashed lines represent the real and imaginary parts of the eigenvalues, respectively. $g=0.1 \omega_c$.}
  \label{fig1}
\end{figure}

To derive the spectrum, we perform a linear transformation, $a=\alpha \cos \theta + \beta e^{-i\Phi/2}\sin \theta$,
$c=-\alpha e^{i\Phi/2}\sin \theta + \beta \cos \theta$, where $\tan 2\theta = - 2ge^{\Phi/2}/(\omega_r-\omega_c)$, and diagonalize the Hamiltonian (\ref{ham})
as $\mathcal{H} =\omega_1 \alpha \alpha^\dagger + \omega_2 \beta \beta^\dagger$, with
the eigenvalues,
\begin{equation}
\omega_{1,2} = \frac{1}{2} \left ( \omega_r + \omega_c \pm 2g\sqrt{ \Delta^2+e^{i\Phi}}\right ),
\label{antic}
\end{equation}
where $\Delta = (\omega_r-\omega_c)/(2g)$ is the detuning.
Figure \ref{fig1} shows a typical spectrum of level attraction ($\Phi=\pi$). Depending on whether the eigenvalues are
real or not, two $\mathcal{PT}$-exact phases when $|\Delta| >1$ and one $\mathcal{PT}$-broken phase when $|\Delta| <1$ seperating by two exceptional points at $|\Delta| =1$ (EP1 and EP2) can be clarified.
Next we will show how magnon and photon interplay to manifest their entanglement properties in these phases.

To proceed, it is essential to know how to describe the hybridized system within the framework of quantum mechanics. In general, the state of the system can be represented by a bi-party density matrix $\rho$. By recasting the effective Hamiltonian as a sum of a Hermitian operator ($\mathcal{H}_H\equiv(\mathcal{H} + \mathcal{H}^\dagger)/2$) and
an antiHermitian operator ($\mathcal{H}_A\equiv(\mathcal{H} - \mathcal{H}^\dagger)/2$), i.e. $\mathcal{H} = \mathcal{H}_H + \mathcal{H}_A$, the dynamic equation of the system
can be expressed as \cite{Brody2012}

\begin{equation}
\frac{\partial \rho}{\partial t}=-i[\mathcal{H}_H,\rho] - i \{\mathcal{H}_A, \rho \} + 2i \mathrm{tr}(\rho \mathcal{H}_A)\rho,
\label{me}
\end{equation}
where the brackets $[~]$ and $\{~\}$ refer to commutator and anticommutator, respectively. We note that the third non-linear term is added to preserve $\mathrm{tr}(\rho)=1$. In general, the purity of a state $\mathrm{tr}(\rho^2)$ is not conserved under Eq. (\ref{me}) unless $\rho^2 =\rho$ because
\begin{equation}
\frac{d \mathrm{tr}(\rho^2)}{dt} =-4i\mathrm{tr}(\rho^2 H_A) + 4i\mathrm{tr}(\rho H_A) \mathrm{tr}(\rho^2).
\end{equation}
This implies that the system will always be locked in a pure state if the initial state is pure ($\rho^2=\rho$).

To solve the evolution of density matrix governed by Eq. (\ref{me}), we start from an initial pure state with mean particle number $N\equiv \langle a^\dagger a + c^\dagger c \rangle =1$. One can immediately prove that
$\partial N/\partial t= 0$ using the commutation relations $[N, H]=0$.
This implies that the particle number is conserved such that the Fock basis $\{|10\rangle$, $|01\rangle \}$ forms
a complete set to describe the system. By solving the eigenequation $H |\phi_k \rangle = \omega_k |\phi_k\rangle$,
we can obtain the eigenstates as,

\begin{equation}
|\phi_k \rangle = \cos \theta_k | 10\rangle + e^{i\varphi_k} \sin \theta_k | 01\rangle,
\end{equation}
where $\theta_k$ and $\varphi_k$ are determined by the relation,
$e^{i\varphi_k}\tan \theta_k = (\omega_k - \omega_r)/g$ ($k=1,2$). Note that the eigenstates are
not mutually orthogonal to each other.

\begin{figure}
  \centering
  \includegraphics[width=0.45\textwidth]{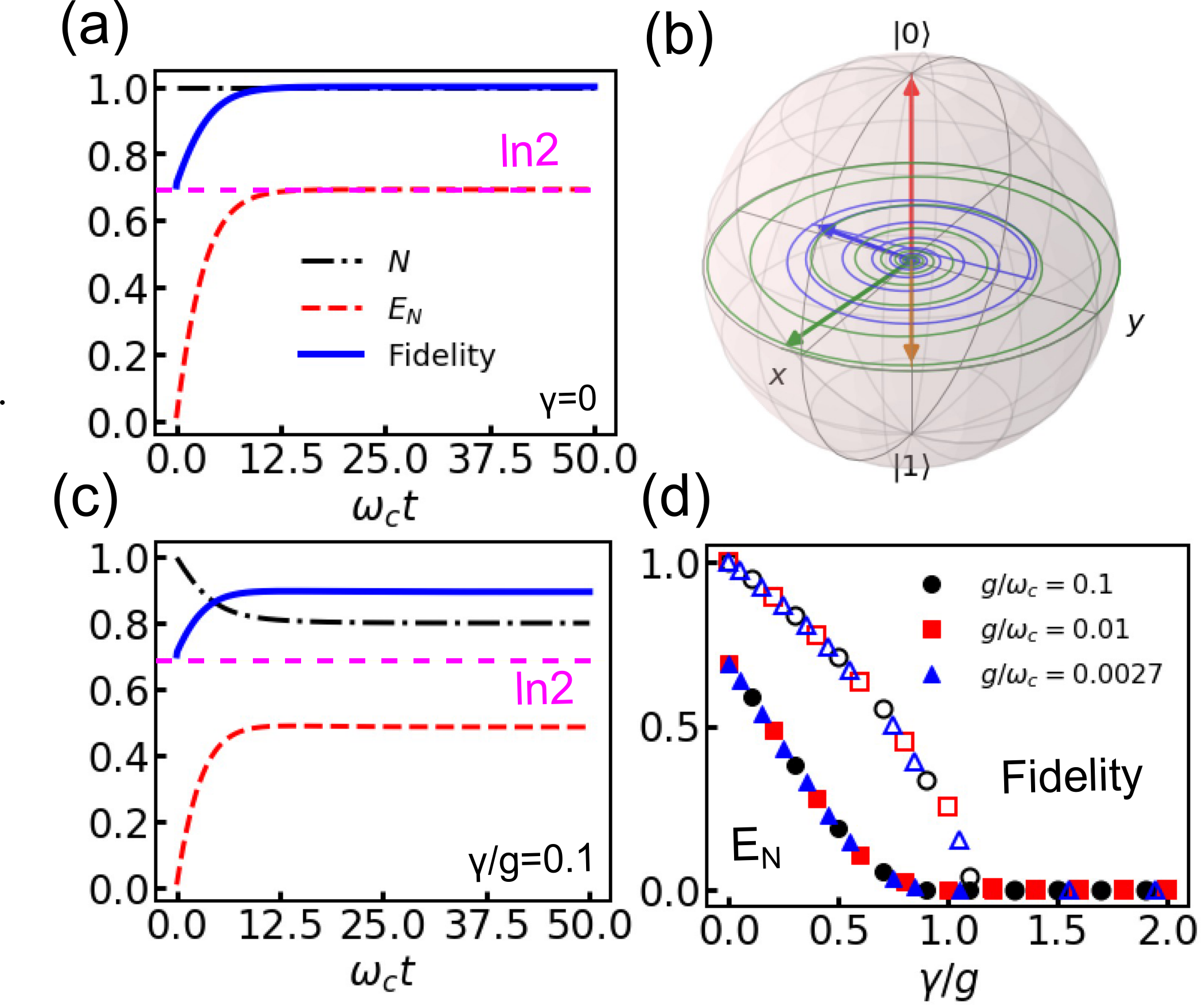}\\
  \caption{(a) Time evolution of particle number density $N$ and entanglement measure $E_N$ under resonance.
  $g=0.1\omega_c, \Delta=0, \rho_0 = |01\rangle \langle 01|$.
  (b) The trajectory of the system in Bloch sphere with initial states $| 01 \rangle \langle 01|$ (red line), $0.6| 01 \rangle \langle 01| + 0.4 \mathbf{I}/4$ (orange line), $| \phi_y \rangle \bigotimes |\phi_y \rangle$ (green line), $ 0.6(| \phi_y \rangle \bigotimes |\phi_y \rangle)( \langle \phi_y | \bigotimes \langle \phi_y |) + 0.4 \mathbf{I}/4$ (blue line),
   where $|\phi_y \rangle=(|1\rangle + | 0\rangle)/\sqrt{2}$ \cite{qutip}.  The Bloch vector $\mathbf{u}$ is obtained by solving
  the equation $(\mathbf{I}+\mathbf{u}\cdot \mathbf{\sigma})/2=\mathrm{tr}_c(\rho)$, where $\mathbf{I}$ is $4\times4$ identity matrix, $\sigma$ is Pauli matrices. The maximally entangled steady state locates at $\mathbf{u}=(0,0,0)$.
  (c) is similar to (a), but with dissipation $\gamma = 0.1g$. The pink dashed line indicates the magnon-photon entanglement
  without dissipation. For simplicity, we choose $\gamma_1=\gamma_2=\gamma$ in the simulations.
  (d) Magnon-photon entanglement and fidelity of the steady state as a function of dissipation in the system.
   $\Delta=0$.}
  \label{fig2}
\end{figure}

Suppose the initial state is $\rho_0 = |01 \rangle \langle 01| =\sum_{k,j} p_{kj} |\phi_k \rangle \langle  \phi_j |$,
then the time-dependent density matrix can be formally written as \cite{Brody2012},

\begin{equation}
\rho = \frac{e^{-i\mathcal{H}t} \rho_0 e^{i\mathcal{H}^\dagger t}}{\mathrm{tr}(e^{-i\mathcal{H} t} \rho_0 e^{i\mathcal{H}^\dagger t})} = \frac{\sum_{k,j} p_{kj} e^{-i\omega_{kj}} |\phi_k \rangle \langle  \phi_j |}{\sum_{k,j} p_{kj} e^{-i\omega_{kj}}\mathrm{tr}(|\phi_k \rangle \langle  \phi_j |)}
\label{rhot}
\end{equation}
where $\omega_{kj} =\omega_{k} - \omega_{j}^*$, and $p_{kj}$ is the expansion coefficients.
 From this density matrix, the magnon-photon entanglement can be quantified through the logarithmic negativity defined as $E_N (\rho) = \ln ||\rho^{T_c}||$, where $\rho^{T_c}$ is partial transpose of $\rho$ with respect to mode $c$ and $||\rho^{T_c} ||$
refers to its trace norm \cite{Vidal2002}. Here $E_N >0$ is a sufficient and necessary condition for magnon-photon entanglement, since the dimension of Hilbert space ($2\times 2=4$) is not larger than six \cite{H1996}. Next we will present the results for energy level attraction and repulsion cases, respectively.

{\it Level attraction.---} For the level attraction case, $\Phi=\pi$. According to the magnitude of detuning
shown in Fig. \ref{fig1}, three regimes can be identified.

(i) $\mathcal{PT}$-broken phase when $|\Delta|<1$. Here both $\omega_1$ and $\omega_2$ are complex numbers and we can derive $\omega_{12}=\omega_{21}=0,\omega_{11}=
-\omega_{22} = 2 ig\sqrt{1-\Delta^2} $ and the expansion coefficients $p_{11}=p_{22}=-p_{12}=-p_{21}=(1-\Delta^2)^{-1}/2$.
As $t \rightarrow \infty$, the $\omega_{11}$ terms dominates both the numerator and denominator of Eq. (\ref{rhot}), such that
the steady density matrix $\rho(t \rightarrow \infty)= |\phi_1 \rangle \langle  \phi_1 |$. Here
$|\phi_{1} \rangle = ( | 10\rangle + e^{i \varphi_1} | 01\rangle)/\sqrt{2}$, where $\varphi_1= \arccos \Delta$ with
the entanglement $E_N(|\phi_1 \rangle \langle  \phi_1 |)=\ln2$. Note that this is a maximally entangled state
for bi-party each with 2-dimensional Hilbert space and it is same as the Bell state
except the neglectable global phase $e^{i \varphi_1}$. Figure \ref{fig2}(a) shows the time evolution of magnon-photon
entanglement (red dashed line) by numerically solving Eq. (\ref{me}), which is consistent with the prediction. To measure the distance between the intermediate state and the steady Bell state, we have introduced the fidelity of the steady state defined as $F( |\phi_1 \rangle,\rho )=
tr\sqrt {\langle  \phi_1 |\rho | \phi_1 \rangle}$ \cite{Nielsen2000}. Clearly, $F$ approaches 1 as the system evolves to the Bell state $|\phi_1 \rangle$ . Figure \ref{fig2}(b) shows the trajectories of
the system in Bloch sphere starting from different initial states represented as red, orange, blue and green arrows, respectively. No matter the initial states are pure (red and green lines) or mixed (orange and blue lines), they all evolve to $|\phi_1 \rangle$ locating at the spherical center. This indicates that $|\phi_1\rangle$
is a fixed point of the system. The underlying physics may be understood as follows: The non-Hermitian nature of the $\mathcal{PT}$-symmetric Hamiltonian results to two eigenmodes with a generic gain ($\omega_1$ mode with positive imaginary component) and loss ($\omega_2$ mode with negative imaginary component).
The particle number in the gain mode will keep increase until all the particles are pumped into this state,
i.e., the system behaves as an attractor to attract all the particles to evolve into the gain mode.

To testify the robustness of this steady Bell state against the perturbation of environment, we introduce the dampings of magnons and photons by adding standard Lindblad operator \cite{Lindblad1976} ($\mathcal{L}\rho= \sum_i\gamma_i (2\xi_i^\dagger \rho \xi_i - \xi_i^\dagger \xi_i \rho - \rho\xi_i^\dagger \xi_i)) $ into the dynamic equation (\ref{me}), where $\xi_{1,2}=a,c$, $\gamma_1$ and $\gamma_2$ are the damping of
magnon mode and photon mode, respectively. By numerically solving the modified dynamic equation, we
obtain the time dependence of magnon-photon entanglement as well as fidelity in Fig. \ref{fig2}(c). Now
the steady state is close but not equal to the Bell state ($F<1$) while the steady entanglement is also smaller than the system without dissipation (pink dashed line). This is expected since the interaction of a quantum system with environment usually
induces the decay of entanglement, with only a few exceptions \cite{yuan20181,yuan20182}. Figure \ref{fig2}(d) shows
the entanglement as a function of dissipation at various coupling strength $g$. Interestingly, the curves with different $g$ perfectly scale in one curve with a critical point at
$\gamma/g=1$, beyond which the entanglement of the steady state disappears. Note that the fidelity of Bell state corresponding to the experimental dissipation ($g/\omega_c=0.0027, \gamma_a=7.6 \times 10^{-5}, \gamma_c=1.5\times 10^{-4}$ \cite{Harder2018}) is $97.85\%$.

\begin{figure}
  \centering
  \includegraphics[width=0.45\textwidth]{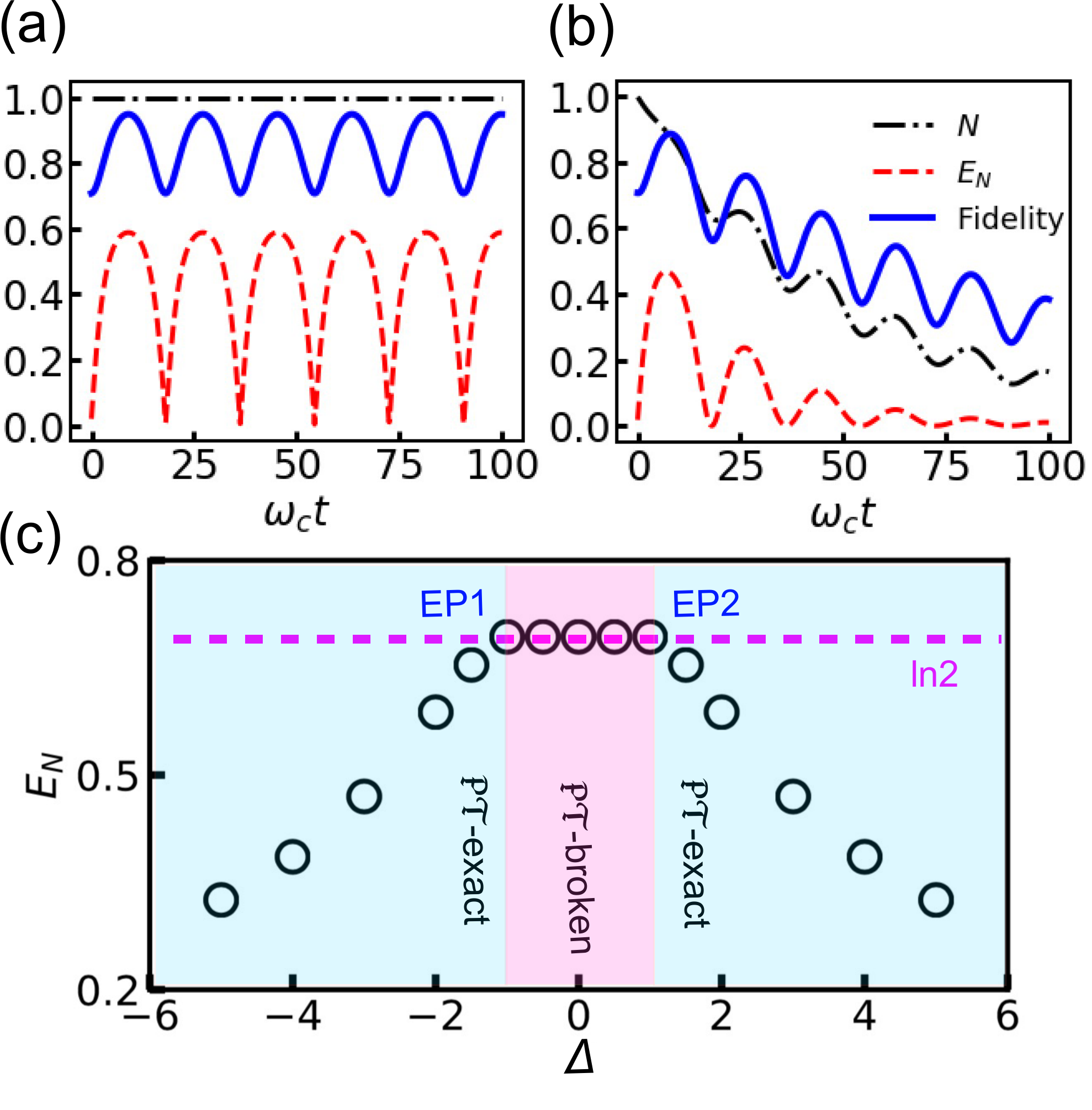}\\
  \caption{(a) Time evolution of particle number density $N$ and entanglement measure $E_N$ away from resonance for $\gamma=0$ (a) and $\gamma=0.1g$ (b), respectively. $g=0.1\omega_c,~ \Delta=2$. (c) Full phase diagram
  of the system with a $\mathcal{PT}$-broken phase sandwiched by two $\mathcal{PT}$-exact phases. The data in the $\mathcal{PT}$-exact phase is taken as the maximum magnon-photon entanglement during Rabi oscillation.}
  \label{fig3}
\end{figure}

(ii) $\mathcal{PT}$-exact phase when $|\Delta|>1$. Here both $\omega_{1,2}$ and $\omega_{ij}$ are real, hence the elements of density matrix
as well as the resulting magnon-photon entanglement will keep oscillating with time, as shown in Fig. \ref{fig3}(a).
No steady state is identified. Once the spontaneous decay of the magnon/photon mode is considered, the particle
number of the system will decay toward zero gradually, accompanied by the oscillating decay of magnon-photon entanglement
towards zero, as shown in Fig. \ref{fig3}(b).

(iii) Exceptional points at $|\Delta|=1$. Here the system has degenerate eigenvalues ($\omega_1=\omega_2=(\omega_r+\omega_c)/2$) and eigenvectors ($|\phi_{1} \rangle = |\phi_{2} \rangle=( | 10\rangle + | 01\rangle)/\sqrt{2}$)
\cite{Miri2019}. At these points, our system will gradually decay into a state with zero entanglement in the presence of dissipation, since no gain exists to resist the the dissipation.

A full phase diagram of the system is shown in Fig. \ref{fig3}(c).
In the $\mathcal{PT}$-broken phase, the magnon-photon entanglement is always steady and maximal ($\ln2$) regardless of the magnitude of detuning.
In the $\mathcal{PT}$-exact phase, the magnon-photon performs Rabi oscillation and thus their entanglement oscillates. The maximum entanglement decreases monotonically as the
detuning increases, which suggests that the steady state is more close to a separable state at large detuning.

{\it Level repulsion case.---} For the level repulsion case, $\Phi=0$, the Hamiltonian is Hermitian
with real eigenvalues such that it is also $\mathcal{PT}$-exact. The resulting magnon-photon
entanglement will keep oscillating with time and no steady entanglement exists \cite{yuan2018apl,Yuan2019}, which is similar to the
$\mathcal{PT}$-exact regime of the level attraction.


\begin{figure}
  \centering
  \includegraphics[width=0.45\textwidth]{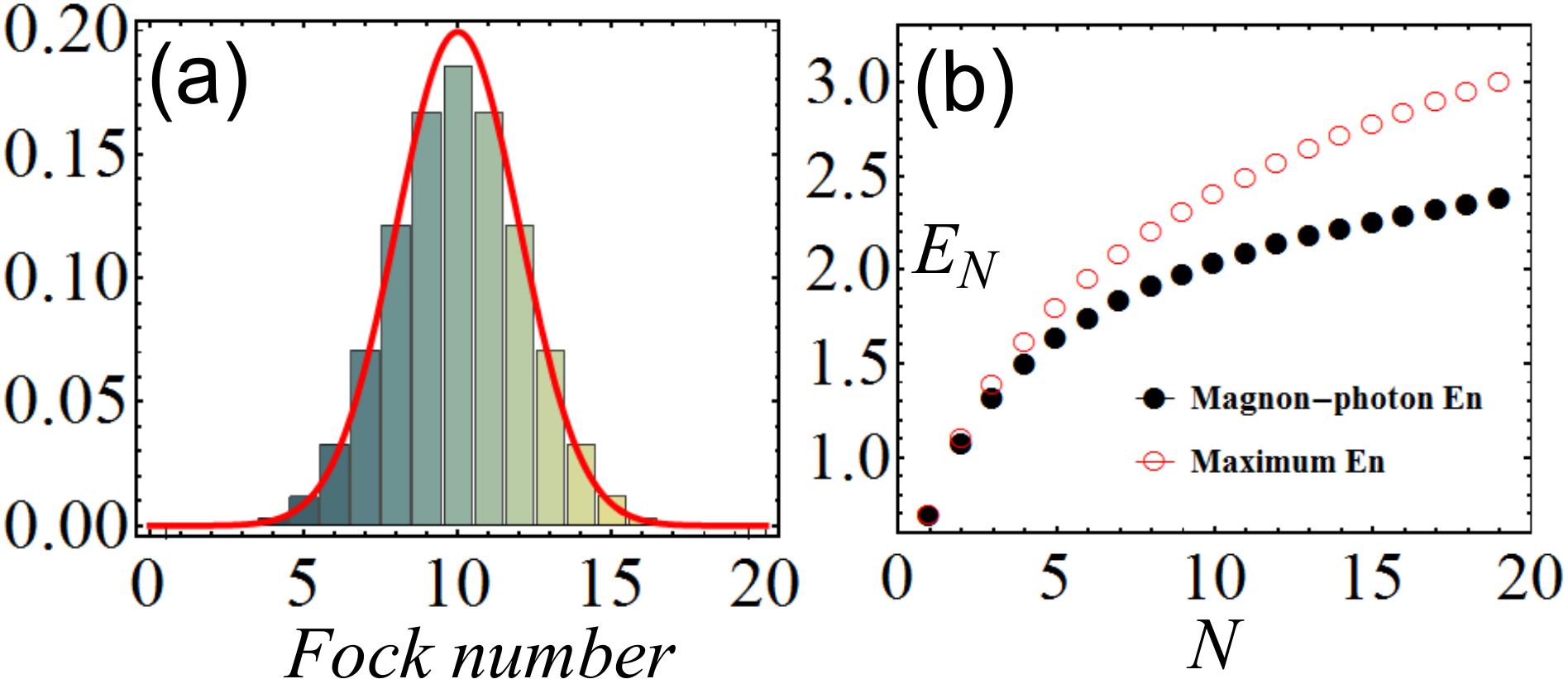}\\
  \caption{(a) Probability distribution of the Fock number of magnons  when $N=19$. The red line is a Gaussian distribution.
  (b) Steady magnon-photon entanglement as a function of total excitation number (black dots) in the $\mathcal{PT}$-broken phase. The red circles represent the maximum bi-party entanglement $E_N=\ln (N+1)$. $\Delta=0$.}
  \label{fig4}
\end{figure}

{\it Discussions and Conclusions.---} We concentrate our previous effort in the Hilbert subspace of $N=1$, if this is not true due to the influence of temperature or large microwave power inside the cavity, the system will tend to reach a multi-particle state
instead of the Bell state. Nevertheless, it is straightforward to solve the Schrodinger equation in Fock space and obtain the steady pure state as $\varphi_=\sum_{k=0}^{ N } c_k | k,N -k\rangle$,
where $k$ and $N -k$ refer to the number of magnons and photons, respectively. Figure \ref{fig4}(a) shows the
distribution of $|c_k|^2$ with respect to the Fock number $k$ when $N = 19$, it approaches a Gaussian distribution
and suggests that the steady state is Gaussian, hence the log-negativity is still a sufficient and necessary condition to quantify the entanglement of magnon and photon \cite{Adesso2007}. Figure \ref{fig4}(b) shows that the entanglement increases with the
particle number (black dots) but the magnitude of entanglement is not larger than the maximally entangled state
$|\varphi \rangle = 1/\sqrt{N +1}\sum_{k=0}^{N} | k,N -k \rangle$ ($E_N=\ln (N+1) $, red circles).
The Bell state generated when $N=1$ is the only point that has maximal entanglement $\ln 2$. This means that
cryogenic environment is essential to generate a Bell state. For commonly used
material YIG, the lowest lying magnon energy is 101 mK \cite{Demotikov2006}, then the typical temperature to excite only one
magnon mode can be estimated from Bose-Einstein distribution as 146 mK, which is accessible in experiments.

Furthermore, the entanglement among magnon, photon, and phonon can also be created through the
nonlinear Kerr effect or magnetostrictive interaction \cite{Li2018,Li2019,Zhang2019}, however, the generated entanglement
is around 0.2 which is much smaller than our finding, probably because of the smallness of nonlinear effect. Note that
all these proposals including ours are constrained to low temperature such that the the excitation of high energy magnons can
be neglected.

In conclusion, we have studied the entanglement properties of magnons and photons inside a cavity and
find that the magnon-photon can form a Bell pair with  maximum entanglement in the $\mathcal{PT}$-broken phase of the system, while no steady entanglement is identified in the $\mathcal{PT}$-exact phase. The generated Bell pair is of high fidelity and robustness against dissipation effect and it is also insensitive to the small detuning between magnon frequency and photon frequency.
 To detect the magnon-photon entanglement, one can measure the particle number fluctuation of the system or perform tomography on the density matrix at low temperature \cite{Bowen2003,Orieux2013}. The generation of the magnon-photon Bell pair
 provides an alternate to achieve maximally entangled state in solid state system and it may be
 utilized as a resource for
 quantum tasks, such as quantum sensing and channel discrimination \cite{Danya2017, Piani2009}.

We acknowledge G. E. W. Bauer, Ka Shen and  V. L. Grigoryan for helpful discussions.
HYY acknowledges the financial support from National Natural Science
Foundation of China (NSFC) Grant (No. 61704071) and Shenzhen Fundamental Subject Research Program
(No. JCYJ20180302174248595). PY was supported by NSFC Grants
(No. 1104041 and 11704060). QH acknowledges NSFC Grants No. 11622428 and No. 61675007, the National Key R\&D Program of China (2016YFA0301302 and 2018YFB1107200).
MHY acknowledges support by Natural Science Foundation of Guangdong Province
(2017B030308003), Guangdong Innovative and Entrepreneurial Research Team Program (2016ZT06D348),
and Science, Technology and Innovation Commission of Shenzhen
Municipality (ZDSYS20170303165926217 and JCYJ20170412152620376).

\clearpage
\onecolumngrid
\end{document}